\documentclass[sigconf]{acmart}
\acmConference[SE 2030]{International Workshop on Software Engineering in 2030}{July 2024}{Puerto Galinàs (Brazil)}
\usepackage{graphicx} 
\usepackage{cleveref}
\usepackage{svg}
\usepackage{titlesec}
\usepackage{enumitem}
\titlespacing*{\section} {0pt}{5ex}{2ex}
\title{Service Colonies: A Novel Architectural Style for Developing Software Systems with Autonomous and Cooperative Services}
\author{Thakshila Dilrukshi}
        \affiliation{%
          \institution{The University of Melbourne}
          \state{Victoria}
          \country{Australia}
          \postcode{3010}
        }
        \email{imiyamohotti@unimelb.edu.au}
        \orcid{0009-0002-7422-2303}

        \author{Artem Polyvyanyy}
        \affiliation{%
          \institution{The University of Melbourne}
          \state{Victoria}
          \country{Australia}
          \postcode{3010}
          }
        \email{artem.polyvyanyy@unimelb.edu.au}
        \orcid{0000-0002-7672-1643}

        \author{Rajkumar Buyya}
        \affiliation{%
          \institution{The University of Melbourne}
          \state{Victoria}
          \country{Australia}
          \postcode{3010}
          }
        \email{rbuyya@unimelb.edu.au}
        \orcid{0000-0001-9754-6496}

        \author{Colin Fidge}
        \affiliation{%
          \institution{Queensland University of Technology}
          \state{Queensland}
          \country{Australia}}
        \email{c.fidge@qut.edu.au}
        \orcid{0000-0002-9410-7217}

        \author{Alistair Barros}
        \affiliation{%
          \institution{Queensland University of Technology}
          \state{Queensland}
          \country{Australia}}
        \email{alistair.barros@qut.edu.au}
        \orcid{0000-0001-8980-6841}
        
\date{March 2024}

\begin{document}

\begin{abstract}
\sloppypar
This paper presents the concept of a \emph{service colony} and its characteristics. 
A \emph{service colony} is a novel architectural style for developing a software system as a group of autonomous software services cooperating to fulfill the objectives of the system.
Each inhabitant service in the colony implements a specific system functionality, collaborates with the other services, and makes proactive decisions that impact its performance and interaction patterns with other inhabitants.
By increasing the level of self-awareness and autonomy available to individual system components, the resulting system is increasingly more decentralized, distributed, flexible, adaptable, distributed, modular, robust, and fault-tolerant.
\end{abstract}

\maketitle

\section{Introduction}

The architecture of a software system defines its structure and behavior in the context of the environment in which the system operates.
Software system architectures evolve to be compatible with changing customer demands.
For example, enterprise application architectures have evolved from legacy mainframe-based software systems and service-oriented architectures to microservices architectures~\cite{NewmanBuildingMicroservices,fowlerLewisMicroservices,serviceOrientedArchitecture}.
Cloud-based technologies accelerate this transition by capitalizing on their scalability, flexibility, security, cost-effectiveness, and monitoring capabilities.
The challenges of modern software systems, however, concern the increasing complexity of managing distributed architectures in highly dynamic environments.
Despite changes occurring in its environment, a software system must continue to operate reliably and efficiently, delivering its functionalities to the users.
Therefore, it is desirable to allow systems to adjust their structures and behaviors at runtime based on the environmental changes to enhance the transparency, elasticity, resilience, and deployment management~\cite{Self-adaptiveMicroservice-basedSystemsLandscapeAndResearchOpportunities2021BergstrMEliaquimPWellingtonP}.

Due to substantial investments already made in software systems, manually migrating them to new architectural styles is often infeasible.
Automated enhancement of existing systems to meet current and future requirements is a more reasonable and cost-effective alternative.
However, ensuring continuous operation under varying conditions often requires extensive human testing and supervision, leading to significant costs for configuring, troubleshooting, and maintaining applications~\cite{KUKUSASelfAdaptiveMicroservicesBanijmaliA2020}.
As a result, there is an increasing demand for automated management and monitoring of distributed systems to reduce costs while ensuring their robustness and quality.

Self-adaptive software systems emerged to reduce uncertainty and complexity in dynamic operating environments~\cite{KUKUSASelfAdaptiveMicroservicesBanijmaliA2020}.
They aim to improve the reliability, availability, flexibility, and performance of systems~\cite{DevelopingSelfAdaptiveMicroserivceSystems2021}.
A self-adaptive autonomous software system monitors its internal and external states, identifying when and how to reconfigure to manage new conditions and dynamically adapt the software architecture at runtime~\cite{DevelopingSelfAdaptiveMicroserivceSystems2021}.
Developing a self-adaptive software system is challenging due to conflicting system goals, the need to monitor different quality of service (QoS) properties, and handling complex mechanisms of adaptation and their effects~\cite{SoftwareEngineeringForSelfAdaptiveSystemsAResearchRoadmap}. 
Once developed, self-adaptive systems can perform self-configuring, self-healing, self-optimizing, and self-protecting functions~\cite{VisionOfAutonomicComputing2003KephartJOChessDM, RainbowArchitectureBasedSelfAdaptationWithReusableInfrastructure2004GarlanDChengS}.

With the increased demand for microservices-based systems, self-adaptive microservices have gained attention.
Even though decentralized decision-making and structural modifications can increase the system's adaptability, the often practiced strategy in this space is a top-down architecture with centralized monitoring and decision-making components~\cite{ComparisonOfApproachesForSelfImprovementinSelfAdaptiveSystems}.
This approach can support the development of agent-based software systems with improved autonomy in individual sub-systems while achieving the overall system's goals~\cite{AnAgentBasedApproachForBuildingComplexSoftwareSystem, OnAgentBasedSoftwareEngineering}.
Existing self-adaptive microservice systems focus on monitoring and managing the adaptation control loop and self-healing properties of the system. 
The prominent adaptation strategy is reactive adaptation, where the adaptation solution applies when the problem occurs~\cite{Self-adaptiveMicroservice-basedSystemsLandscapeAndResearchOpportunities2021BergstrMEliaquimPWellingtonP}.
Structural changes in the application layer via service aggregations and splittings, as well as the introduction and removal of communication links between system services, would increase the level of flexibility of self-adaptive systems, providing scope for autonomous adaptations of the system's architecture.
However, instead of the application layer, the focus in self-adaptive microservices is often on the management of the infrastructure layer, aiming to optimize system performance.
Additionally, aspects relevant to the requirements of edge computing and executing systems under resource-constrained environments~\cite{MicroSplitRahmanianA2022} are essential characteristics of future software systems to enhance the quality of the services provided to end users.

This paper introduces \emph{service colonies}, a software architectural style for developing systems as groups of autonomous software services that cooperate to fulfill the global objectives of the overall system, increasing the level of autonomy and intelligence of the distributed (micro)services-based systems.
Each inhabitant service of a colony fulfills a specific part of the functionality of the overall system the colony implements.
By following prescribed or acquired through learning rules and by interacting with other inhabitants, the services demonstrate a swarm-like global intelligence in adapting the individual services, their deployment in the environment, and the configuration of communication links between the services and the environment that ensures continuous effective and efficient performance of the system.
Though individual services can have different roles, no system components are envisioned to implement centralized control over the behavior and configuration of the overall system.
An inhabitant service of a colony can interact with other services and the environment.
Through interactions of individual services with the environment, the system interacts with its users to receive tasks and return results, as well as the hardware, software, and network infrastructure in which the system is deployed to optimize the utilization of the available resources.
The interactions between the services ensure complex functions composed of functionalities of the interacting services can be realized.

By monitoring their behavior and environment, the inhabitant services of a colony identify their own and the overall system’s performance bottlenecks and resource constraints and automatically adapt themselves and their communication links to overcome these issues. 
Therefore, service colonies follow a proactive adaptation strategy where the actions are taken before problems occur.
Consequently, a service colony monitors, adapts, and optimizes its behavior through self-diagnosis and adaptations of individual components based on environmental conditions. 
Thus, it self-manages and self-architects in a dynamic environment.
Moreover, it is an autonomous, goal-oriented, and goal-directed system that aims to enhance the quality of service provided to its users.
In this way, service colonies extend self-adapting software systems, aiming to reduce the costly human interventions typically required for continuous monitoring, troubleshooting, and application maintenance.

The remainder of the paper proceeds as follows.
\Cref{sec:BackgroundAndRelatedWork} discusses existing service-based software architectures.
Then, \Cref{sec:Guidelines} presents service colonies.
\Cref{sec:BenitifsOfServiceColony} discusses the envisaged benefits of this new architectural style, while \Cref{sec:challenges} identifies challenges associated with the development of systems as service colonies.
Finally, \Cref{sec:Conclusion} concludes the paper.

\section{Related Work}
\label{sec:BackgroundAndRelatedWork}

In this section, we discuss three areas:
(micro)service-oriented architectural styles~\cite{serviceOrientedArchitecture, fowlerLewisMicroservices, NewmanBuildingMicroservices},
self-adaptive software systems~\cite{TheEvlovingPhilosophersProblemDynamicChangeManagement1990KramerJMageeJ, SelfManageSystemsAnArchitecturalChallenge2007KramerJMageeJ}, and 
self-adaptive (micro)services-based systems~\cite{AnAutomatedApproachtoManagementofaCollectionofAutonomicSystems2019, DynamicSelf-AdaptationForDistributedServiceOrientedTransactions2012GommaHHasimotoK}.

Service-oriented architectures~\cite{serviceOrientedArchitecture} and microservices~\cite{fowlerLewisMicroservices, NewmanBuildingMicroservices} are software architectural styles that describe a system as a group of communicating services.
Often, such a system identifies a primary service responsible for communications within the system.
When a system receives a request, it delegates the necessary functions to its constituent services.
While executing a specific functionality, a service can communicate with other services in the system.
The response to the user is provided by aggregating the results of potentially multiple service calls.

\begin{figure} [t]
\vspace{2mm}
\centering
\includegraphics[scale=0.3, trim = 30mm 45mm 30mm 15mm]{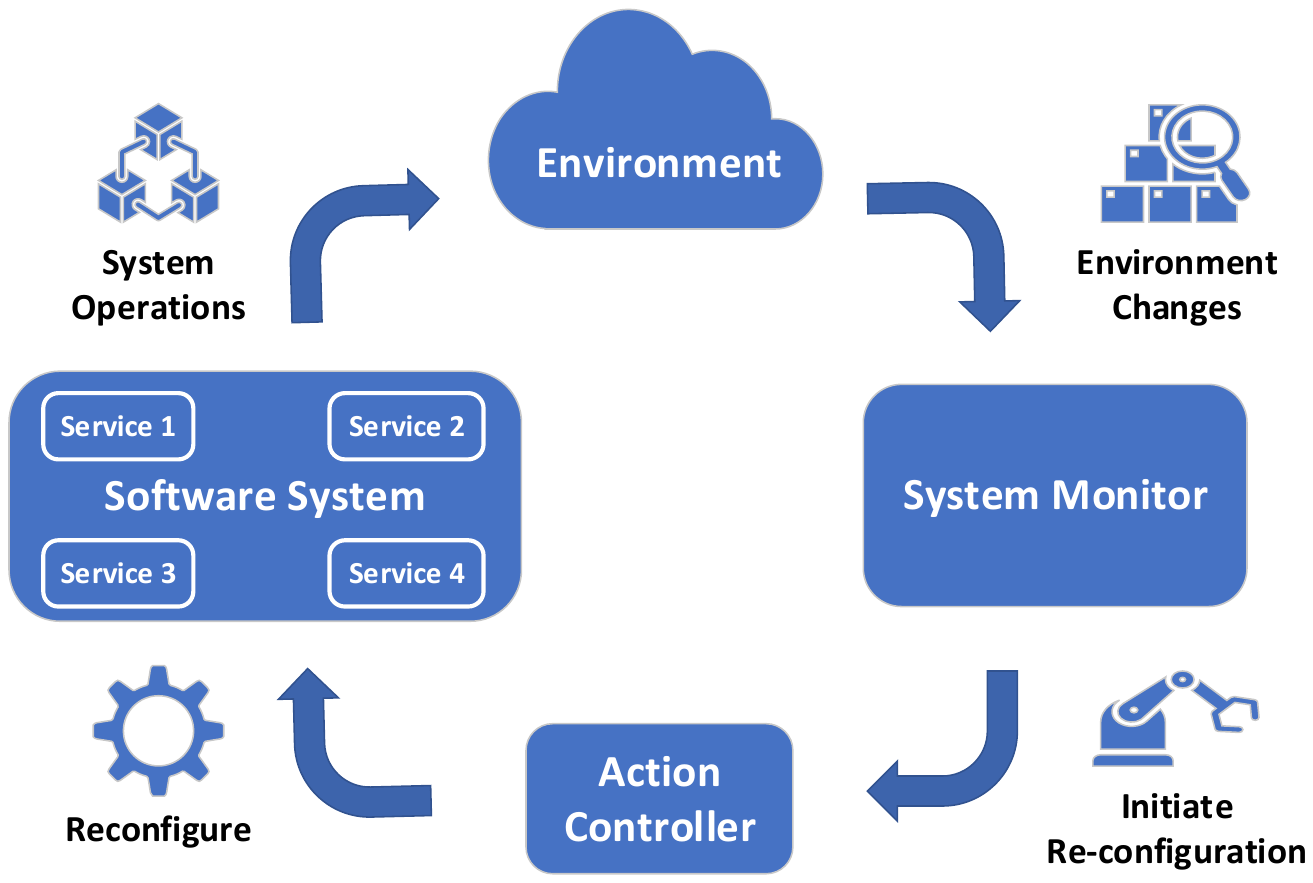}
\caption{Self-adaptive system architecture}
\label{fig:autonomousSystem}
\vspace{-5mm}
\end{figure}

A self-adaptive system architecture is illustrated schematically in \Cref{fig:autonomousSystem}.
A self-adaptive system can learn by observing the environment in which it is embedded.
It has monitoring and decision-making components that are illustrated in \Cref{fig:autonomousSystem} as the system monitor component and the action controller component, respectively.
The monitor component observes the environmental changes~\cite{RealizingSelfAdaptiveSystemsViaOnlineReinforcementLearning2022MetzgerA, DevelopingSelfAdaptiveMicroserivceSystems2021, ApplyArchitectureBasedAdaptation2018WeynsD}.
Once a significant change is identified, the monitor triggers the reconfiguration request.
The action controller component identifies the required changes based on static rules or dynamically identified interventions.
Subsequently, the action controller can initiate the reconfiguration of the system.
Finally, the reconfigured system is deployed in the environment.
Such self-adaptations are implemented as a repetitive, ongoing process.

The prominent technique for developing self-adaptive software systems is through control loop~\cite{SoftwareEngineeringForSelfAdaptiveSystemsAResearchRoadmap}, which overviews the main activities of the system with a feedback cycle that starts with the relevant data collected from the system environment.
The MAPE-K control loop~\cite{AnArchitecturalBlueprintForAutonomicComputing2006DavidS, VisionOfAutonomicComputing2003KephartJOChessDM} monitors, analyzes, plans, and executes self-adaptive software systems~\cite{ApplyingMachineLearningInSelfAdaptiveSystems2021GheibiOWeynsDQuinF, ApplyingMAPE-Kcontrolloopsforadaptiveworkflowmanagemeninsmartfactories2023MalburgL, RealizingSelfAdaptiveSystemsViaOnlineReinforcementLearning2022MetzgerA}.
Policy-based adaptations implement a static approach, where system experts analyze and design a policy repository for condition-triggered adaptations~\cite{PolicyBasedSelfAdaptiveSystems2010NargesKRamtinKMarjanSSaeedJ,SoftwareArchitectureBasedSelfAdaptation2009GarlanDSchmerlBChengS}.
These include pre-defined scenarios and corresponding actions to adapt the system.
Rainbow~\cite{RainbowArchitectureBasedSelfAdaptationWithReusableInfrastructure2004GarlanDChengS}, SASSY~\cite{SASSYAFrameworkForSelfArchitectingServiceOrientedSystems}, MUSIC~\cite{MUSICADevelopmentFrameworkAndMethodologyForSelfAdaptingApplications}, and REACT~\cite{REACTAModelBasedRunTimeEnvironmentForAdaptiveCommunicationSystems} are example frameworks for software systems with self-adaptation control loops.
However, these frameworks are not tailored towards decentralized systems of communicating services and require a profound knowledge of self-adaptive systems development~\cite{REACTAModelBasedRunTimeEnvironmentForAdaptiveCommunicationSystems}.
Rainbow specifies an abstract architecture model to validate runtime properties of the system, evaluate them against the model of constraint violations, and perform global- and module-level adaptations of the system~\cite{RainbowArchitectureBasedSelfAdaptationWithReusableInfrastructure2004GarlanDChengS}.
SASSY (Self-architecting Software Systems) is a model-driven framework that, based on pre-defined software adaptation patterns, aims to accommodate requirement changes in a dynamic environment.
It generates candidate software architectures and selects the one that best serves stakeholder-defined, scenario-based quality-of-service (QoS) goals~\cite{SASSYAFrameworkForSelfArchitectingServiceOrientedSystems}.
Finally, MUSIC provides a model-driven development approach combined with middleware that facilitates dynamic and automatic adaptation based on adaptation concerns~\cite{MUSICADevelopmentFrameworkAndMethodologyForSelfAdaptingApplications}.
SASSY, MUSIC, and REACT are all based on the MAPE-K control loop principles.

Self-adaptation in SOA has been discussed in the literature~\cite{DynamicSelf-AdaptationForDistributedServiceOrientedTransactions2012GommaHHasimotoK, Cross-layerSelfAdaptattionOfService-OrientedArchitectures}.
Existing works propose middleware that supports cross-layer adaptation of SOA systems by considering the server-side perspectives~\cite{Cross-layerSelfAdaptattionOfService-OrientedArchitectures}.
This approach supports the adaptation of the service interface and application layer.
It is based on a common meta-model of the two layers.
Furthermore, dynamic self-adaptation in distributed service-oriented transactions~\cite{DynamicSelf-AdaptationForDistributedServiceOrientedTransactions2012GommaHHasimotoK} is an extension of the SASSY framework~\cite{SASSYAFrameworkForSelfArchitectingServiceOrientedSystems}.
This extension proposes dynamic software adaptations in distributed transactions using a two-phase commit protocol.
Specifically, it defines adaptation patterns and state machine models for transaction commits.

An automated approach for managing a collection of autonomic systems is based on the concept of a meta-manager that uses a parameterized adaptation policy~\cite{CaseStudyOfAnAutomatedApproachToanagngCollectionsOfAutonomicSystems2020ThomasJGarlanDSchemerB, AnAutomatedApproachtoManagementofaCollectionofAutonomicSystems2019}.
In this approach, a system is seen as composed of multiple constituent systems.
Each constituent system contains a manager that keeps track of system metrics directly related to the overall system's quality of service objectives.
Each manager selects the actions that will improve the system's performance.
The approach implements a hierarchical control strategy that uses control theory to manage the behavior of the system while outsourcing decision-making responsibility to individual meta-manager units.
However, this approach does not address the problem of structural changes in the application layer.
Instead, it performs resource/capacity optimizations based on the runtime cost and response time analysis.

Based on the systematic mapping study~\cite{Self-adaptiveMicroservice-basedSystemsLandscapeAndResearchOpportunities2021BergstrMEliaquimPWellingtonP}, the state-of-the-art self-adaptive microservices-based systems focus on infrastructure layer adaptations~\cite{ImprovingMicroserviceBasedApplicationsWithRuntimePlacementAdaptation2019, ModelBasedGenerationOfSelfAdaptiveCloudServices2019, ASelfHealingMicroservicesArchitecture2020} or multi-layer adaptations~\cite{TowardsReferenceArchitectureForAMultilayerControlledSelfadaptiveMicroserviceSystem2018, Cross-layerSelfAdaptattionOfService-OrientedArchitectures, AMultiLevelSelfAdaptationApproachForMicroserviceSystems2019}, while no studies address the problem of self-adaptation strategies in the application layer.
Furthermore, there are works on reactive adaptation strategies, in which adaptations are applied after problems are identified~\cite{ImprovingMicroserviceBasedApplicationsWithRuntimePlacementAdaptation2019, ModelBasedGenerationOfSelfAdaptiveCloudServices2019, ASelfHealingMicroservicesArchitecture2020}.
Proactive adaptation strategies have been proposed for scientific workflows~\cite{MONAD2017} and IoT architectures~\cite{DataDrivenAdaptationInMicroservicebasedIoTArchitectures2020}.
Centralized monitoring has been practiced as the prominent adaptation control mechanism~\cite{ImprovingMicroserviceBasedApplicationsWithRuntimePlacementAdaptation2019, TowardsReferenceArchitectureForAMultilayerControlledSelfadaptiveMicroserviceSystem2018, AMultiLevelSelfAdaptationApproachForMicroserviceSystems2019}.

The state-of-the-art studies on self-adaptive systems focus on the areas of cloud-based services, such as IoT and IaaS, rather than service-based software systems~\cite{SelfAdaptiveSystemsASystematicLiteratureReview2022TerenceWMarkusWChrisstophT}.
These approaches, which span across various domains (robotic, IoT, communication, automotive, e-commerce), utilize centralized monitoring, pre-defined models, and rule-based techniques~\cite{SelfAdaptiveSystemsASystematicLiteratureReview2022TerenceWMarkusWChrisstophT, ComparisonOfApproachesForSelfImprovementinSelfAdaptiveSystems}.
The introduction of self-managing and self-architecting principles in software systems could increase the sustainability of (micro)service-based architectures.

\section{Service Colonies}
\label{sec:Guidelines}

A \emph{service colony} is a system composed of software services, or \emph{inhabitants}, that interact with each other to perform specific functions.
The inhabitants are deployed in a distributed computing environment, such as a cloud, edge computing infrastructure, or IoT devices.
Being intrinsically distributed, a service colony can exercise different levels of scatter in its control and decision-making capabilities across individual inhabitants, ranging from decentralized via divisional or hierarchical to centralized.

An \emph{inhabitant} of a colony is responsible for performing some dedicated functionality, a \emph{service}, within the system and exhibits a degree of autonomy.
Hence, inhabitants within a colony may have diverse characteristics and capabilities, reflecting their different roles and responsibilities.
Colony inhabitants have goals they aim to achieve and strive to optimize their performance and the performance of the overall system.
Therefore, they are self- and situationally-aware entities.
Inhabitants monitor their performance relative to their responsibilities, accepted service obligations, and the environment, and plan actions that ensure the desired quality of service.
In doing so, they can proactively make decisions and take actions, such as replicating, migrating to another, more performant compute node, splitting into two entities, and seeking integration with another colony's inhabitants.
An inhabitant may rely on learning mechanisms to improve their operations and decision rules based on prior experiences, allowing them to adapt and enhance their behavior.

An inhabitant can interact with other inhabitants and the environment.
An inhabitant can request services from other inhabitants or provide service to them.
In general, interaction patterns in a service colony can range from simple message exchanges between two inhabitants to complex collaboration protocols involving multiple inhabitants and the environment.
Through these interactions, inhabitants coordinate activities, exchange information, and take inputs from and return outputs of the system to the users.

The environment of a service colony comprises its inhabitants, the computing infrastructure utilized by the colony, and the communication channels that the colony inhabitants can exploit.
As special components in the environment, we identify the users of the overall colony system.
The users seek to accomplish tasks and functions by interacting with interface entities of the colony and receiving results of such requests.
The users seek to accomplish tasks and functions by interacting with interface entities of the colony and receiving the results of such requests.
The environment of a service colony is, thus, inherently dynamic and stochastic.

\begin{figure}[tb]
\vspace{2mm}
\centering
\includegraphics[scale=0.3, trim=20mm 40mm 50mm 20mm]{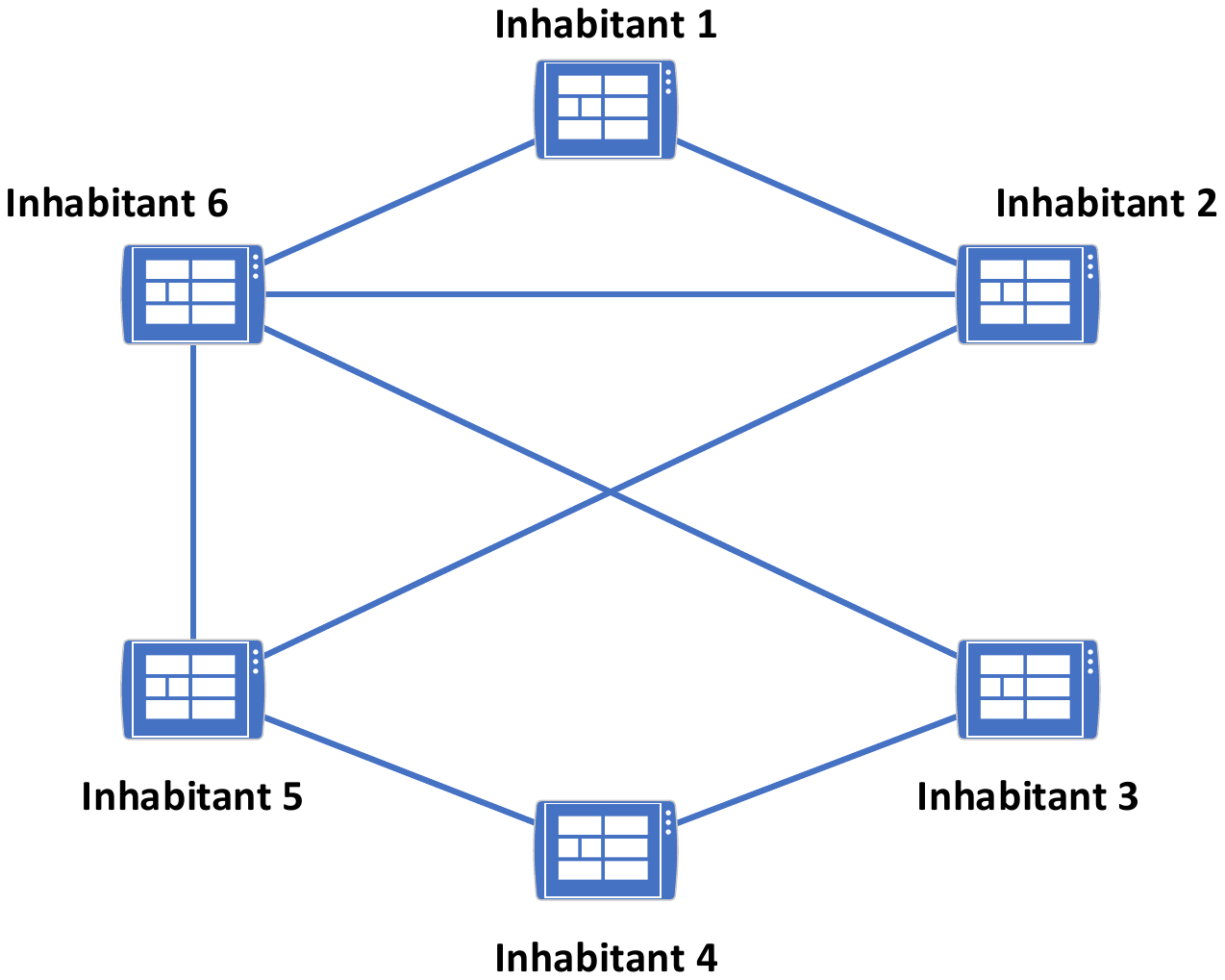}
\caption{An example service colony architecture}
\label{fig:systemArchitecture}
\vspace{-5mm}
\end{figure}

As opposed to a self-adaptive system, a service colony has decentralized monitoring.
In this proposed architecture, all the inhabitants act as autonomous agents.
There is no special inhabitant responsible for overall communication, monitoring, or dynamic adaptations.
Instead, each inhabitant can learn from its behavior and initiate actions to adjust the system automatically.
Each individual inhabitant of a service colony can be (sub-)optimal, but collectively, they aim to optimize the overall system performance.
\Cref{fig:systemArchitecture} sketches an example composition of service inhabitants in a service colony.
Arcs between inhabitants indicate the links used to support communications between the individuals.
Inhabitants can communicate with other inhabitants, for example, via messages passing through communication links.
Analytical messages and behavioral messages are the two types of special messages sent to the environment.
Each inhabitant can initiate a change in the colony based on its analysis of itself and the environment using analytical messages.
These can involve an inhabitant splitting, merging with another inhabitant, or adding or removing communication links between inhabitants.
Furthermore, an inhabitant can send behavioral messages to the environment, indicating its capabilities or limitations.
Examples of behavioral messages include the availability of more resources to be occupied by another inhabitant, delays in requests or responses, or the volume of data in requests or responses.
 
\subsection{Inhabitant Characteristics}

An inhabitant is the core building element of a service colony system.
It encapsulates a delegated service, part of the functionality of the overall system, and monitors and proactively optimizes its performance.
It is a self-contained entity.
Each inhabitant has a boundary and a designated functionality that can be clearly distinguished from other inhabitants.
Inhabitants are autonomous agents that can act and react independently.
Inhabitants have a dynamic state that can change over time.
Inhabitants have a state that can change over time.
The future actions taken by an inhabitant depend on its current state.
Inhabitants can share their state with other inhabitants in the colony.
It is intended that an inhabitant frequently communicates and delivers services to other inhabitants.
Thus, the current state of an inhabitant can be influenced by the state of another inhabitant in the colony and its behavior.
Inhabitants are exploratory, and they can learn and adapt to the environment.
They can learn from the environment and adapt their behavior based on their experience.
Therefore, inhabitants can proactively adjust the system based on their behavior.
Each inhabitant has goals, making it goal-directed and goal-oriented.
An inhabitant attempts to accomplish these goals via its behavior and optimize the individual and overall system objective via adaptive learning.

\subsection{Inhabitant Architecture}

\begin{figure}[tb]
\vspace{1.4mm}
\centering
\includegraphics[scale=0.28, trim=0mm 25mm 0mm 24mm]{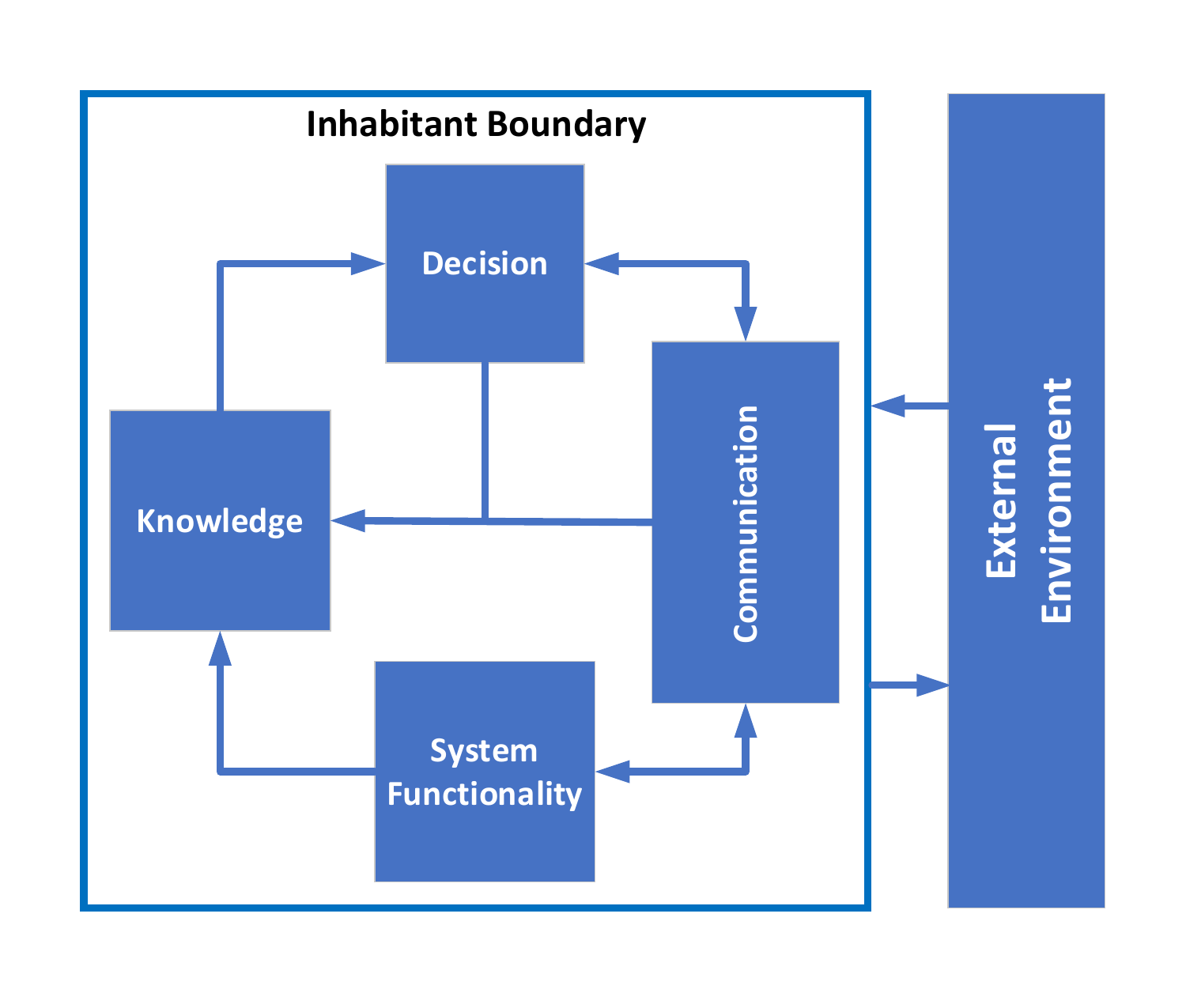}
\caption{Inhabitant architecture}
\label{fig:agentArchitecture}
\vspace{-4mm}
\end{figure}

\Cref{fig:agentArchitecture} depicts the architecture of an inhabitant.
Each inhabitant has a boundary and can send messages to and receive messages from the external environment. 
The sub-modules of an inhabitant include system functionality, communication, knowledge, and decision-making.
These sub-modules communicate based on their responsibilities to provide the functionalities of the inhabitant.

The \emph{communication} sub-module manages all interactions of the inhabitant and serves as its entry and exit point.
Messages received from the environment, including those from other inhabitants, are captured by the communication module and forwarded to the relevant sub-modules.
Messages initiated by the inhabitant and addressed to the environment are also routed through the communication sub-module for dissemination to their recipients.
Upon receiving a message, the communication module validates its content to determine which sub-module is responsible for further processing.
For example, functionality-related messages, such as requests to perform a function, are directed to the system functionality sub-module.
Messages concerning environmental behavior are forwarded to the knowledge sub-module, while those pertaining to configurations and adaptations of inhabitants are redirected to the decision sub-module.
Messages sent to the environment are routed to their intended recipients.
Analytical messages can target individual inhabitants, subsets of inhabitants, or all inhabitants within the service colony.
Behavioral messages are delivered to colony inhabitants based on the communication links within the topology of the colony.


The service colony has obligatory system functionalities that are distributed across its inhabitants.
The \emph{system functionality} sub-module is responsible for executing the inhabitant's delegated functionalities.
Each inhabitant collaborates with other inhabitants in the service colony to fulfill these obligations.
After completing a delegated function, an inhabitant responds to the inhabitants or users in the environment who have requested the result of the function via the communication module.
Additionally, it sends log data to the knowledge sub-module to collect internal behavioral information.


Inhabitants monitor themselves and their environment.
The \emph{knowledge} sub-module builds the inhabitant’s intelligence that aggregates the experiences of its monitoring processes.
During initialization, an inhabitant collects data related to the service colony’s structure and behavior.
This information includes the total number of inhabitants, their configurations, and initial behavioral data of inhabitants.
During execution, data and experiences stem from two sources.
Functional data is collected from the system functionality sub-module.
Additionally, an inhabitant gathers behavioral and analytical data of the service colony via the communication sub-module.
The inhabitant does not accept all the data received from the environment.
It filters data based on quality and relevance.
Furthermore, it tracks the origin of the data for the decision-making process executed by the decision sub-module.


Inhabitants can react to changes in their environment.
They learn and take action based on their behavior and the evolution of the environment.
The \emph{decision} sub-module handles these learning and decision-making functions.
This sub-module depends on data provided by the knowledge sub-module.
Learning is driven by data available in the knowledge sub-module and previous decisions made by the decision sub-module.
An inhabitant can use adaptive learning to dynamically adjust to the environment based on the current and past behavior of the system.
Data collected after implementing a decision is used as feedback for learning.
The decision sub-module handles two functions.
First, if data from the inhabitant indicates a relevant behavior, it disseminates that message to the environment via the communication sub-module.
These are behavioral messages.
Second, it analyzes inhabitant and environment data to identify system operations, such as operations relevant to rearchitecting the system, and actions for improving performance.
Such information is shared with other inhabitants in the colony via analytical messages.
Each inhabitant has objectives captured as collections of rules within the decision sub-module.
These rules can be updated dynamically via the interface provided to the decision sub-module.
Therefore, during execution, new rules can be added, while existing rules can be modified or deleted.

\subsection{Operations}

An inhabitant can replicate itself. 
If an inhabitant experiences a high volume of requests and no further optimizations of the inhabitant are possible, it can decide to replicate itself to accommodate the high volume of service requests.

\begin{figure}[tb]
\centering
\includegraphics[scale=0.3, trim=30mm 30mm 30mm 20mm]{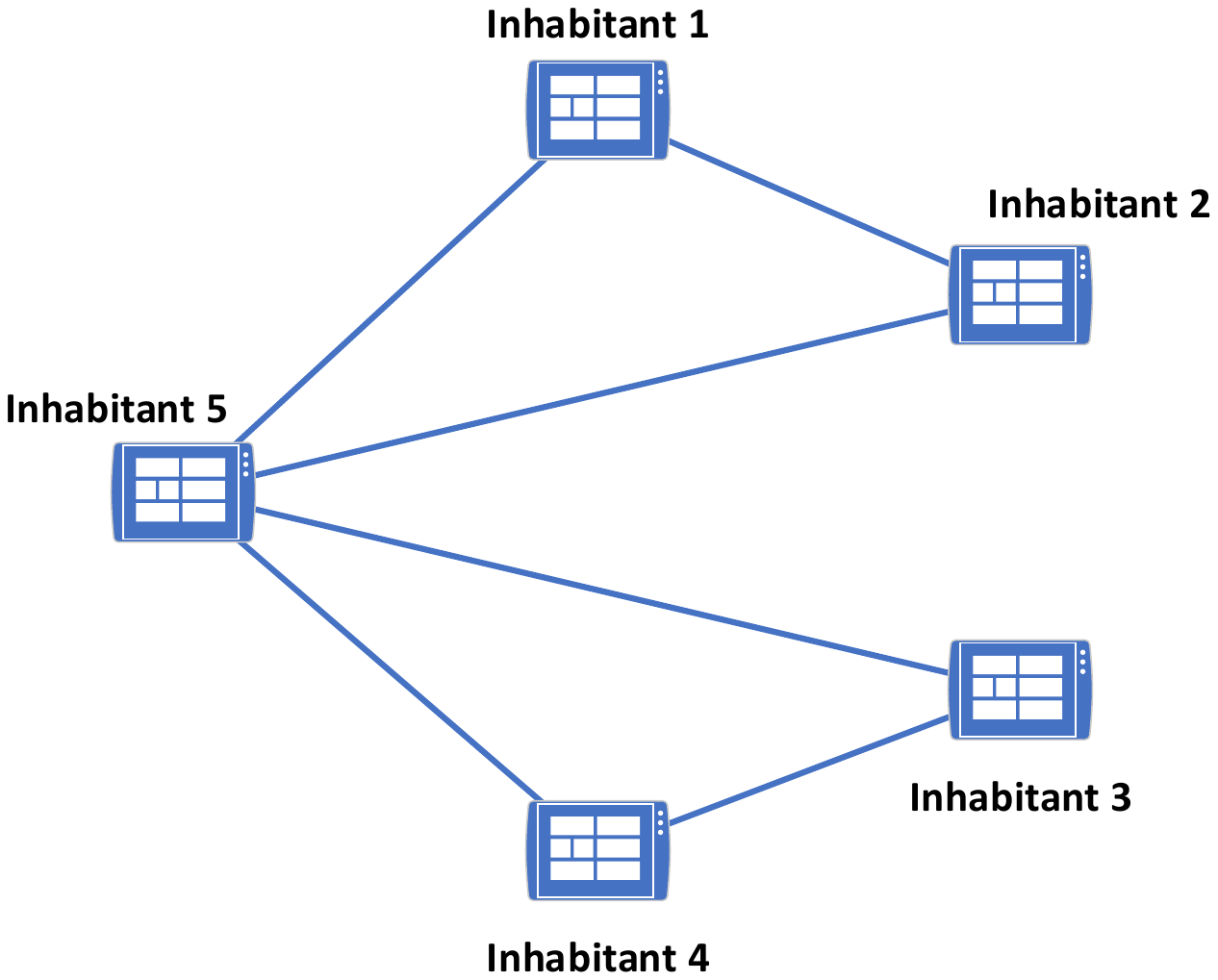}
\caption{Service colony after merging}
\label{fig:inhabitantMerging}
\vspace{-4mm}
\end{figure}

A service colony can change its structure to adapt to the environment and optimize its communication patterns.
Four basic operations are proposed to support such adaptations: joining two or more inhabitants, splitting an inhabitant into two or more inhabitants, and adding and removing communication links between inhabitants.
Two or more inhabitants can \emph{join} to form a single inhabitant, for example, to simplify the communication structure between themselves and the rest of the colony during a low system load.
Alternatively, such joining can be triggered to maximize the resource utilization in the system, hence reducing the operation cost.
A complex join of multiple inhabitants can be implemented as a sequence of atomic joins between pairs of inhabitants.
That is, two inhabitants merge in the first step. 
Then, another inhabitant joins with the previously merged inhabitant.
An inhabitant can \emph{split} into multiple inhabitants via a sequence of atomic splits of one inhabitant into two inhabitants to distribute its functionality and responsibilities among multiple system elements.
Such splittings can be initiated based on the identified performance bottlenecks and resource-intensive functionalities that degrade the quality of the service provided.
Hence, it can split its functionalities to maximize performance, for instance, by replicating those decomposed services that receive high request loads.
Finally, an inhabitant can establish direct communication links with other inhabitants, for example, to reduce current communication latency or to identify peer inhabitants from which to request required services.
Moreover, merging and splitting operations in the system could lead to the addition and removal of communication links in the colony.

\begin{figure}[tb]
\centering
\includegraphics[scale=0.3, trim=30mm 30mm 30mm 20mm]{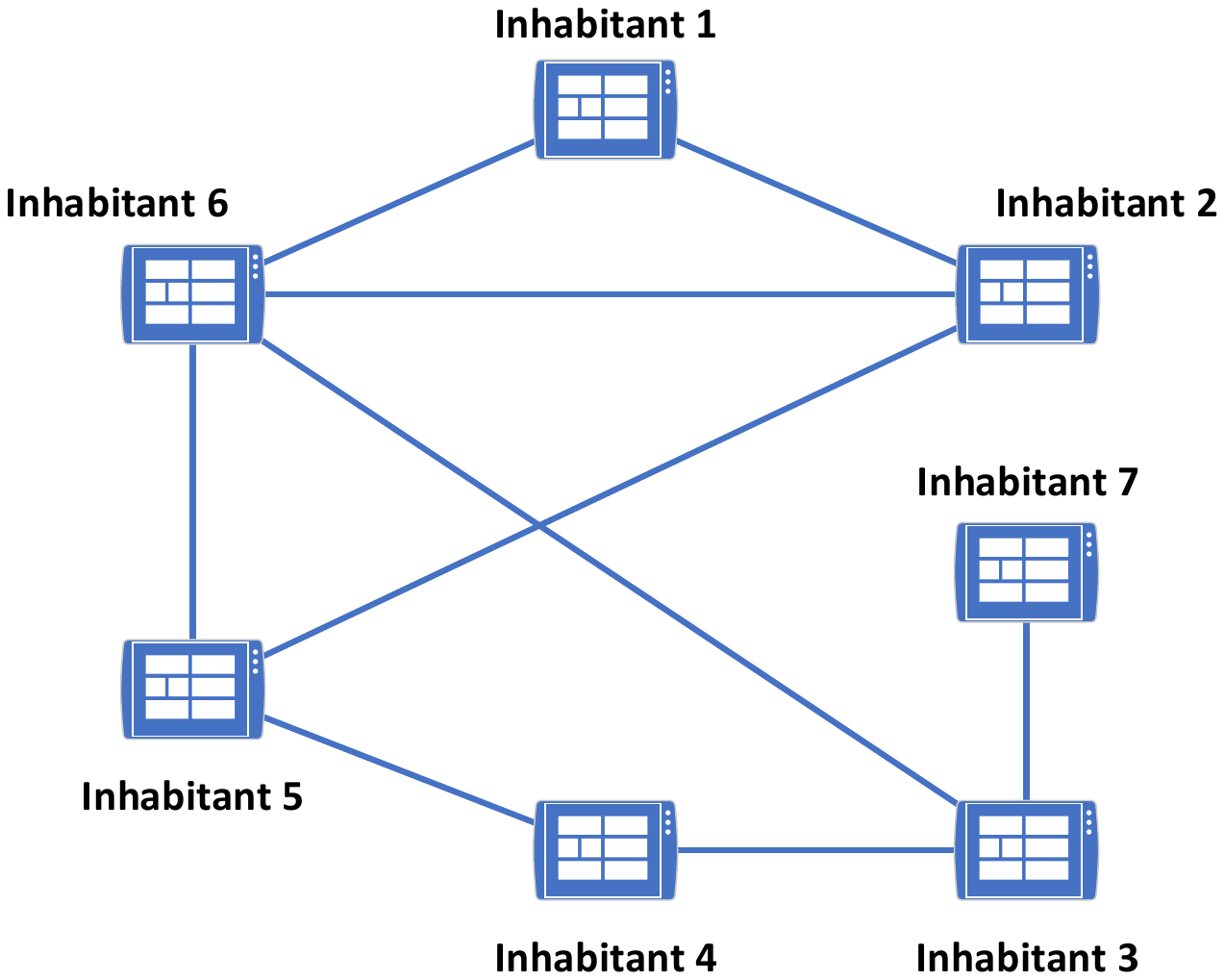}
\caption{Service colony after splitting}
\label{fig:inhabitantSplitting}
\vspace{-4mm}
\end{figure}

\Cref{fig:inhabitantMerging} depicts the re-architectured system from \Cref{fig:systemArchitecture} after merging inhabitant~6 with inhabitant~5.
This operation leads to redirecting the communication links of inhabitant 6 to and from the resulting inhabitant~5. 
An example splitting of inhabitant~3 is depicted in \Cref{fig:inhabitantSplitting}.
In this case, inhabitant~7 is split out of the original inhabitant~3 in the system.
In this example, a communication link is established between the resulting inhabitants. 
The inhabitants can use this link to request services from each other. 

One can come up with different strategies for deciding which adaptations of the system structure to perform.
Such strategies can result in the authoritative execution of the intended adaptation or an adaptation after confirmation from peers, confirmation from a group of inhabitants, or confirmation from the entire service colony.
Inhabitants execute authoritative actions without requesting confirmation from other members of the colony.
Alternatively, an inhabitant can negotiate the intended splittings and joins with other inhabitants to ensure mutual agreement and maximal benefit for all the negotiators.

Inhabitants can request confirmation from their peers in case the change affects them.
Similarly, if the change affects a group of inhabitants of the colony, permissions can be requested from the impacted inhabitants.
For instance, if inhabitant~6 from \Cref{fig:inhabitantSplitting} decides to split into two services, it may require obtaining permission from inhabitants~1, 2, 3, and~5 since they have direct communication links with inhabitant~6.
Execution of an adaptation decision is initiated by sending a message to the relevant inhabitants to request permission.
Once these requests arrive at the recipients, they validate the requested change.
The validation is based on the knowledge stored within the decision sub-modules.
Once validated, the inhabitants reply with approvals or rejections of the change.
If the change is approved, the requesting inhabitant executes the change process.
At the beginning of this process, the inhabitant confirms the start of the change to the environment.
After the inhabitant receives a confirmation to start the process, it should not accept further adaptation requests until the current execution process is completed and confirmed.
Then, the inhabitant executes the change. 
After completing the change, it informs the environment by sending the execution completion message.
Then, other inhabitants update their configurations based on the executed change.
These updates can involve adding, removing, or updating new or existing links between inhabitants.

\begin{figure}[t]
\vspace{-5mm}
\centering
\includegraphics[scale=0.5]{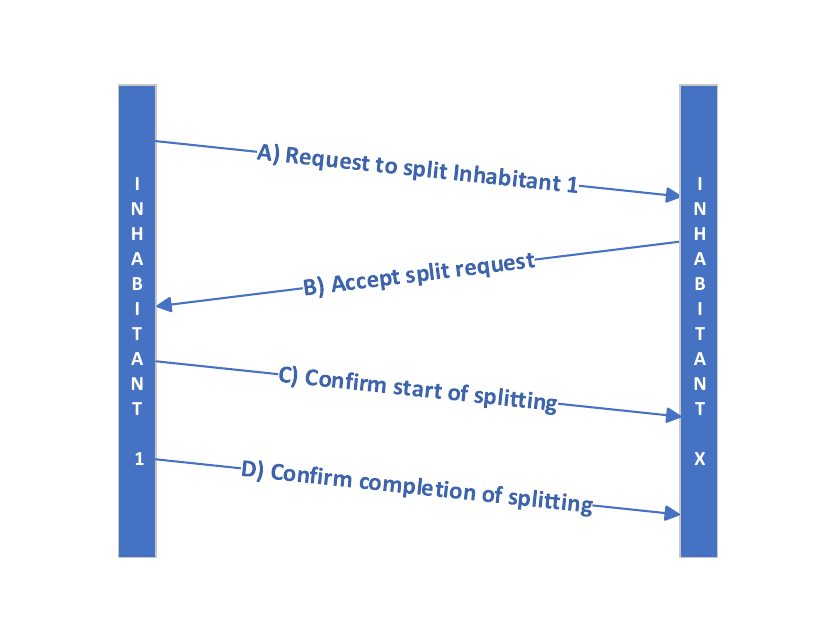}
\vspace{-8mm}
\caption{Splitting request communication sequence}
\label{fig:splittingAcceptRequest}
\vspace{-6mm}
\end{figure}

\begin{figure}[ht]
\vspace{-1mm}
\centering
\includegraphics[scale=0.34]{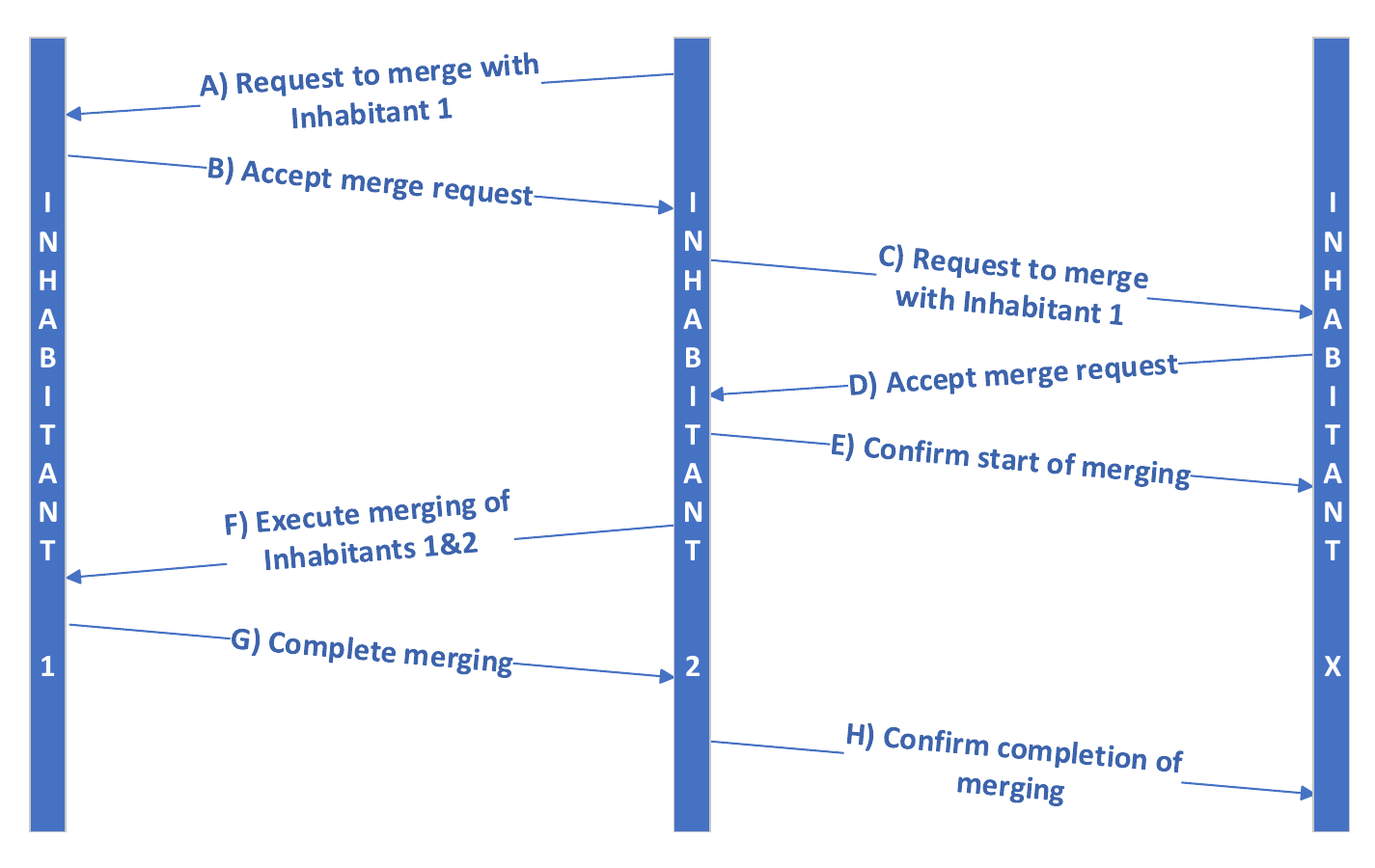}
\vspace{-2mm}
\caption{Merging request communication sequence}
\label{fig:MergeAcceptRequest}
\vspace{-5mm}
\end{figure}

\Cref{fig:splittingAcceptRequest} illustrates an example communication sequence for splitting an inhabitant in a service colony.
An inhabitant requests the environment to split.
After other inhabitants accept the request, the splitting task is confirmed, and the splitting process is executed.
After successful splitting, the confirmation regarding the successful splitting is sent to the environment.
Then, the relevant inhabitants of the colony update their knowledge and links. 
If the inhabitant’s request to split is rejected, no further actions are taken, and the communication ends.
\Cref{fig:MergeAcceptRequest} illustrates the scenario of merging two inhabitants.
In this case, inhabitant~2 identifies the need to merge and acts as the leading inhabitant of the change process. 
First, it seeks to get confirmation from inhabitant~1, the inhabitant it intends to merge with. 
Once inhabitant~1 accepts the merging request, the next request from inhabitant~2 to the environment is sent to get approval to merge.
If the environment responds positively, the merging process commences.
After the merging process is completed, the execution confirmation message is sent to the environment to update relevant configurations.

\section{Benefits}
\label{sec:BenitifsOfServiceColony}

Both service-oriented architecture and microservices architecture identify loosely coupled services that can be deployed separately from the main application.
These individual services can be independently designed, developed, and tested.
Furthermore, using containerization techniques, services can auto-scale in a cloud-based environment based on their performance. 
In general, a self-adapting system handles specific domain requirements and resource optimizations and adaptations in the infrastructure layer using reactive error-handling functionality executed in pre-modeled situations~\cite{SelfAdaptiveSystemsASystematicLiteratureReview2022TerenceWMarkusWChrisstophT, ComparisonOfApproachesForSelfImprovementinSelfAdaptiveSystems}.
Service colonies, however, are not restricted by pre-identified or pre-modeled scenarios. 
A colony can proactively re-architect itself using a bottom-up approach and decentralized analysis of the system.

Consider an online shopping system developed as a service-based application.
Assume that user registration and shopping cart actions are developed as two separate services within the system.
In December, due to the festive season, the system receives a 100\% surge in the volume of requests.
During this high workload period, shopping cart services can experience performance bottlenecks, for example, due to latency in the payment handling process.
A standard way to address this scenario is to increase resources for the entire shopping cart actions service.
If this system is implemented as a service colony, it can identify that the bottleneck is caused by the payment transactions and split out this functionality into a new service.
Subsequently, the colony can aim to utilize available resources to scale the payment transactions, for instance, by replicating the new payment service while keeping other shopping cart actions within the original service.
These nuanced adaptations can drastically reduce the operating costs of the system.
Moreover, when the system experiences a low volume of requests, the shopping cart actions and payment handling services can merge back to achieve a smaller system footprint.
Note that within a service colony, such splitting and merging happen automatically, reducing the costs of maintaining the system.

The benefits of a service colony are, thus, at least the following:

\begin{itemize}[leftmargin=*]
\item
\textbf{Extension of Microservices.}
Service colonies build upon the strengths of microservice-based architectures, inheriting their flexibility, resilience, modularity, optimized resource usage, and reduced operational overhead. Since the industry is moving toward broader adoption of microservice-based systems, service colonies contribute to this trend, providing more options for engineering future software systems.
\item
\textbf{Enhanced Flexibility.}
By fostering interactions among individual inhabitants, a service colony ensures that desired functionalities are delivered flexibly. 
For example, multiple inhabitants can deliver the same functionality based on different service agreement levels or with improved performance by proactively scaling or migrating the functionality to more productive compute nodes.
In addition, each inhabitant can be developed, deployed, tested, and updated independently.
\item
\textbf{Proactive Fault Tolerance.}
Each inhabitant in the service colony generates its own analytics and shares them throughout the colony.
The individual and collaborative analysis of this data supports effective predictions of the system’s behavior, including potential future faults.
Since the analysis is based on the runtime information of individual services, it can be used to effectively monitor, predict, and proactively respond to environmental changes and system faults.
\item
\textbf{Continuous Optimization.} 
Through continuous learning from its operations, each inhabitant, and consequently the entire service colony, evolves into a self-optimizing entity, constantly refining itself for enhanced performance.
\item
\textbf{Goal Orientation.} 
Both the service colony as a whole and its individual inhabitants have objectives and strive to achieve them. 
The collaborative nature of the system fosters iterative improvements toward shared goals.
\item
\textbf{Dynamic Service Introduction and Composition.} 
The dynamic introduction and integration of services within a service colony optimize system performance and resource utilization, seamlessly adapting to evolving demands and environmental changes.
\item
\textbf{Heterogeneity.}
A service colony hosts diverse inhabitants, each potentially utilizing different technologies and implementations.
However, adherence to standard communication protocols ensures effective data dissemination throughout the ecosystem.
\item
\textbf{Scalability.}
By facilitating individual scalability based on performance metrics, a service colony can reduce maintenance costs through proactive decision-making and predictive scaling initiated by its inhabitants.
\end{itemize}

\section{Challenges}
\label{sec:challenges}

The design, implementation, and maintenance of a service colony come with challenges that are yet to be understood and studied.
We initiate this endeavor by discussing several such challenges below:

\begin{itemize}[leftmargin=*]
\item
\textbf{Complex System.}
A service colony is a dynamic, complex system composed of interacting components.
Without hierarchical control or global coordination, in general, slight modifications of components and interaction patterns can have a substantial impact on the overall high-level behavior of the system.
Without hierarchical control or global coordination, even slight modifications to components and interaction patterns can significantly impact the system's overall behavior.
This intricate relationship between low-level component behaviors and high-level system behavior complicates error identification and troubleshooting.
Additionally, if the splitting of inhabitants is not controlled, the colony can proliferate excessively, increasing system latency.
Although the system aims to optimize for latency, unnecessary adjustments can result in the opposite effect.
\item 
\textbf{Verification and Validation.}
To verify the correctness of a system, it is essential to test it under different conditions, for example, by simulating environmental changes. 
The distributed nature of a service colony complicates this process due to the vast number of possible scenarios the system can execute.
New tools are required to simulate different workloads and environmental changes to properly test service colonies.
Additionally, validating the correctness of system decisions before rearchitecting is challenging due to the system's dynamic nature, which makes it difficult to predict outcomes and ensure stability in advance.
\item 
\textbf{Dynamic Updates.} 
A service colony is a dynamic system.
Adjustments within such a system can lead to temporary unavailability of certain functionalities.
Therefore, sophisticated mechanisms are needed to manage these dynamic adjustments effectively, ensuring smooth operations during system rearchitecting.
\item
\textbf{Heterogeneity.} 
A service colony can host diverse inhabitants, each implemented using different technologies.
To support this heterogeneity, standard interfaces and communication protocols must be established.
Moreover, the integration of components implemented using different technologies complicates system development, testing, and monitoring.
Furthermore, such a heterogeneous system is more vulnerable to security threats. 
\item
\textbf{Persistence.} 
A service-based system relies on databases and data caching layers for information storage.
These components require non-trivial adaptations during the reengineering of the system.
Dynamic repartitioning and redesigning of databases can lead to data replication, which may result in data inconsistencies.
\item 
\textbf{System Updates and Change Requests.}
System updates are mandatory to comply with industry standards, and customer change requests are inevitable. 
Implementing these changes in a distributed system is a complex task.
A robust system state management process must be in place to handle updates effectively.
Automatically persisting the system state before and after adjustments is essential for maintaining system operations.
Additionally, a service colony must be equipped with comprehensive mechanisms for the automatic deployment of services resulting from the splitting of colony inhabitants and for managing continuous delivery and integration pipelines.
\item 
\textbf{Reengineering.} 
Existing systems that wish to benefit from the advantages of service colony architecture need to be reengineered accordingly. 
A systematic process for reengineering software systems into service colonies must be defined to facilitate the migration of legacy systems to this new architectural style.
\end{itemize}
\section{Conclusions}
\label{sec:Conclusion}

This paper introduces the concept of a \emph{service colony}, a software architectural style for developing systems as groups of autonomous, interacting software services.
Each inhabitant service in a colony is driven by its aim to deliver services to its users, either external users of the system or other inhabitants of the colony.
Based on their past performance, inhabitants can proactively decide to self-replicate, split into multiple services, or join with other inhabitants to ensure high-quality service delivery.
In this way, the overall service colony system can adapt to changing workloads by either shrinking its footprint during periods of low workload or scaling specific high-demand functionality during workload bursts.
By performing such adaptations, the system aims to minimize resource utilization while maximizing the quality of the delivered services over time.
A service colony is a bottom-up complex system characterized by numerous interacting components that result in emergent system-level behavior.
Consequently, service colonies aim to inherit the benefits of complex system architectures, including resilience, robustness, adaptability, scalability, and distributed control.
Future work on service colonies will focus on designing and evaluating different colony inhabitant architectures and principles of their interactions to study their effects on the global behavior of the overall system.

\smallskip
\noindent
\textbf{Acknowledgements.}
This work was in part supported by the Australian Research Council project DP220101516.

\bibliographystyle{ACM-Reference-Format}
\bibliography{main} \label{References}
\end{document}